\documentclass{paper}

\usepackage{amsmath}
\usepackage{amssymb}
\usepackage{graphicx}
\usepackage{epsfig}
\usepackage{cancel}

\newcommand{\acceleration}{\boldsymbol{a}}
\newcommand{\velocity}{\boldsymbol{v}}

\newcommand{\Dfrac}{\displaystyle\frac}

\begin{document}

\title{From Kepler's laws  to Newton's law: a  didactical proof}

\author{Riccardo Borghi\\
Dipartimento di Ingegneria\\
Universit\`a degli Studi ``Roma tre''\\
Via Vito Volterra 62, I-00146 Rome, Italy\\
{\tt riccardo.borghi@uniroma3.it}
}

\maketitle

\begin{abstract}
An elementary derivation of the
 Newton ``inverse square law''
from the three Kepler laws  is  proposed. Our proof, thought  essentially 
for first-year  undergraduates,  basically
rests on Euclidean geometry.  It could then be offered even
to high-school students possessing only the first basics of Calculus.
\end{abstract}


\newpage\

\begin{quotation}
{\em
\noindent What makes planets go around the sun? At the time of Kepler some people answered this problem by saying that there were angels behind them beating their wings and pushing the planets around an orbit. As you will see, the answer is not very far from the truth. The only difference is that the angels sit in a different direction and their wings push inwards. }
\end{quotation}

(Richard P. Feynman~\cite{Feynman/1965})

\newpage

\section{Introduction}
\label{Sec:Introduction}

It is generally easier to find mathematical derivations of 
Kepler's laws from Newton's inverse square law (Newton's law henceforth) rather
that the opposite, the probably most famous one being that given by Richard Feynman in his 
celebrated ``Lost Lecture''~\cite{Goodstein/Goodstein/1970}.
However, the importance of the fact that Newton's law is a logical consequence of Kepler's laws
was  considerably  emphasized by Max Born in his beautiful textbook on cause and chance~\cite{Born/1948}.
Born poses such a logical consequence ``... as the basis on which my (his) whole conception of 
causality in physics rests,''~\cite[p.~129]{Born/1948}, and furnished   a full rigorous proof 
by using the most natural coordinate system for dealing with problems involving 
central  forces, namely the polar one~\cite[Appendix~II]{Born/1948}.  
In 1993 a compact and interesting mathematical derivation of Newton's law from Kepler within Cartesian 
realm has been published in~\cite{Hyman/1993} and, to use the author own words, ``... without need of
the `clever tricks' that are often used when polar coordinates are employed.''

The aim of the present paper is purely pedagogical.  
In fact, although  Kepler's laws and Newton's law are central topics in any first-year 
undergraduate physics course,  the mathematical  background and knowledge of the audience 
is still too far from being acceptable for a complete presentation to be adequately grasped.
As far as my teaching experience is concerned, this implies that a rigorous justification of Newton's law
is carried out only for the simplest case of circular orbits while it is left unsolved for elliptical orbits.
Such an unsatisfactory state of fact pushed me to conceive a proof to be offered also to first-year  
undergraduates, or even to high-school students possessing only the first basics of Calculus.  

To help teachers, the present work is organized in the form of a self-contained didactical unit, which
can be provided in three steps. In fact, since our proof is ultimately based on two geometrical properties of 
ellipses, which could not necessarily be known to students, a couple of appendixes have been added to the 
paper. Each appendix could then constitute the subject of a practice session.

\section{The proof}
\label{Sec:Proof}

\subsection{Kepler's laws}
\label{Subsec:KeplerLaws}

Here the Kepler laws are listed for reader's convenience:

\begin{itemize}

\item[I.] Each planet moves along an ellipse with the Sun at one of the  two foci;

\item[II.] The segment joining the Sun and the planet sweeps out equal areas in equal times;

\item[III.] The square of the orbital period divided by the cube of the elliptical orbit major axis is the same for all planets. 

\end{itemize}

The proof then follows within the following three steps.

\subsection{Step 1: use of first Kepler's law}

\begin{figure}[!ht]
\centerline{\includegraphics[width=6cm,angle=90]{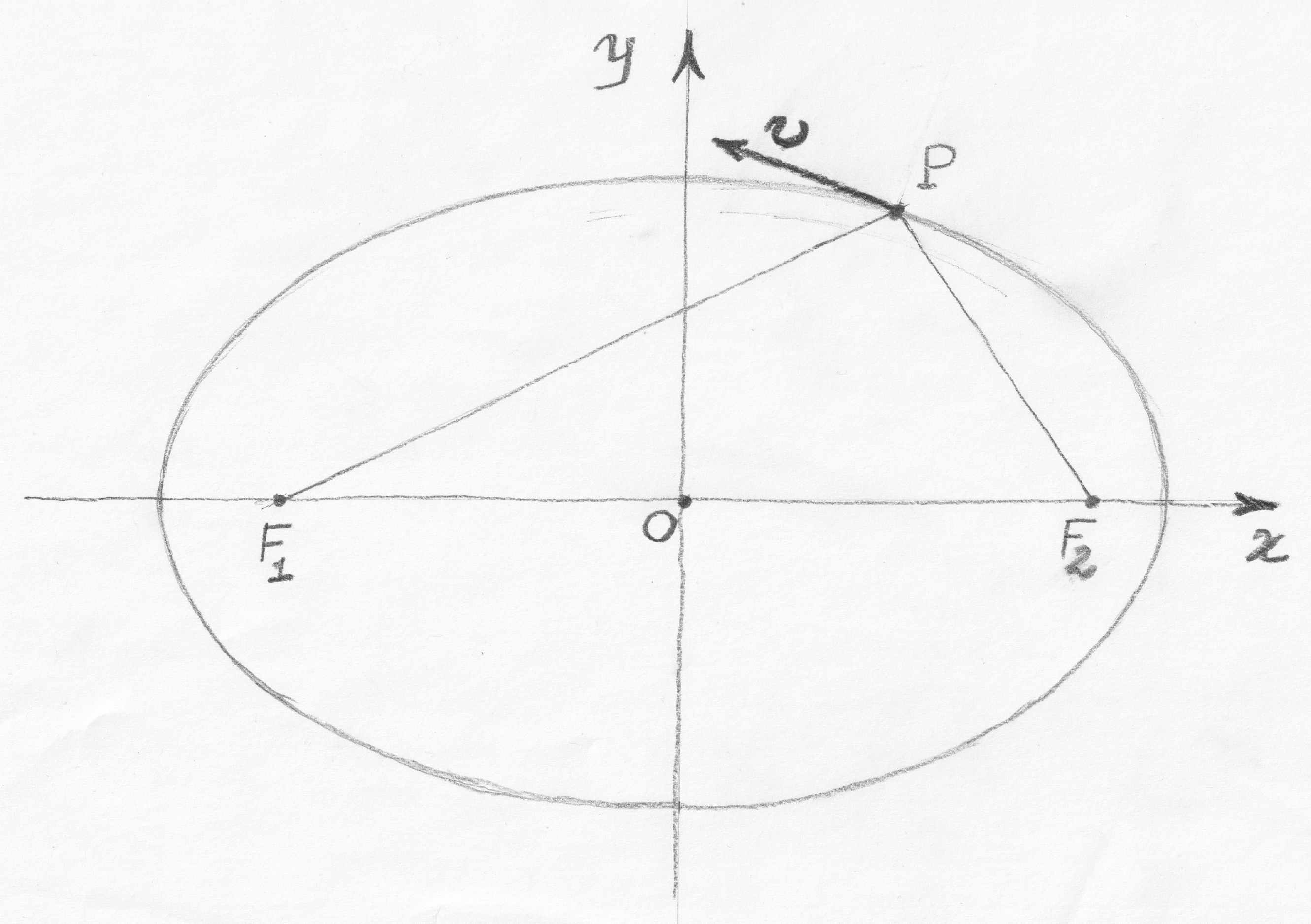}}
\caption{The first Kepler law.} 
\label{Fig:Geometry.1}
\end{figure}

We start from the  Kepler law I and  the geometry is depicted in Fig.~\ref{Fig:Geometry.1}:
 the planet is represented by the point $P$ which is supposed to  move counterclockwise with velocity 
 $\velocity$ along 
the elliptical trajectory whose foci are  $F_1$ and $F_2$. The Sun is at $F_1$. 
We have 
\begin{equation}
\label{Eq:EquazioneGeometricaEllisse}
\overline{PF_1}\,+\,\overline{PF_2}\,=\,2a\,,
\end{equation}
where $2a$ denotes the ellipse major axis. 
Needless to say, in the following of the paper
the symbol $a$ should not  be confused with the modulus of the 
acceleration $\acceleration$, which will then be denoted by 
$|\acceleration |$. Moreover, the symbol $2f$ stands for the focal distance  $\overline{F_1F_2}$,
in such a way the orbit eccentricity, say $\epsilon$, is given by the ratio $f/a$ and the minor half-axis, 
say $b$,  turns out to be
\begin{equation}
\label{Eq:EquazioneGeometricaEllisse.0}
b\,=\,\sqrt{a^2\,-\,f^2}\,=\,a\,\sqrt{1\,-\,\epsilon^2}\,.
\end{equation}
Moreover, a Cartesian reference frame has also been introduced in such a way the ellipse representation 
reads
\begin{equation}
\label{Eq:EquazioneCartesianaEllisse}
\Dfrac {x^2}{a^2}\,+\,\Dfrac {y^2}{b^2}\,=\,1\,.
\end{equation}
The first of the above quoted properties of the ellipse we are going to use within our proof is
the following:
\begin{quotation}
{\em P1: the normal to the ellipse at $P$ bisects the angle $\sphericalangle F_1 P F_2$ }
\end{quotation}
A clever, purely geometrical proof of P1 can be found, for instance, in the above quoted 
Feynman ``Lost Lecture''~\cite[pp.~150-151]{Goodstein/Goodstein/1970}. 
A  less clever, although straightforward proof, based on the Cartesian 
representation~(\ref{Eq:EquazioneCartesianaEllisse}), 
is outlined in~Appendix~\ref{App:A} and could be used, as it was said above, as the topic of a 
preparatory practice session. 

\subsection{Step 2: use of second Kepler's law}
\label{Subsec:KeplerLawII}

In the following it will be assumed that all students know that the areal speed, which will be denoted 
$\dot{\mathcal{A}}$, can be  mathematically defined in terms of the  cross product between  the position 
vector $\overrightarrow{F_1P}$ and the point velocity   $\velocity$ as
\begin{equation}
\label{Eq:Areal.0}
\dot{\mathcal{A}}\,=\,\Dfrac 12\,\overrightarrow{F_1P}\,\times\,\velocity\,.
\end{equation}
Moreover, since the motion is planar, in the following the  symbol $\times$
has to be meant as the sole cross product component along the direction perpendicular to the
motion plane. In other terms, it is thought of as a \emph{scalar} quantity.
Kepler's law II  then implies  that the areal speed $\dot{\mathcal{A}}$  is a constant of the motion.
This, in turns, has two consequences. 

The first of them is that the acceleration of $P$ points toward $F_1$. 
The original geometrical proof provided by Newton has been summarized 
again in~\cite[pp.~153-157]{Goodstein/Goodstein/1970}.
An alternative way is to take the time derivative of both sides of Eq.~(\ref{Eq:Areal.0}), and 
on taking into account that, since the areal speed is constant, it turns out that $\ddot{\mathcal{A}}=0$, 
so that
\begin{equation}
\label{Eq:Areal.1}
2\ddot{\mathcal{A}}\,=\,\cancel{\dot{\overrightarrow{F_1P}}\,\times\,\velocity}\,+\,
\overrightarrow{F_1P}\,\times\,\dot{\velocity}\,=\,\overrightarrow{F_1P}\,\times\,{\acceleration}\,=\,0\,,
\end{equation}
where use has been made of the fact that $\dot{\overrightarrow{F_1P}}=\velocity$ and that
$\dot{\velocity}=\acceleration$. Equation~(\ref{Eq:Areal.0}) then implies that ${\acceleration}\,\parallel\,\overrightarrow{F_1P}$.

The second consequence of Kepler's law II represents the central point of our proof.
\begin{figure}[!ht]
\centerline{\includegraphics[width=6cm,angle=-90]{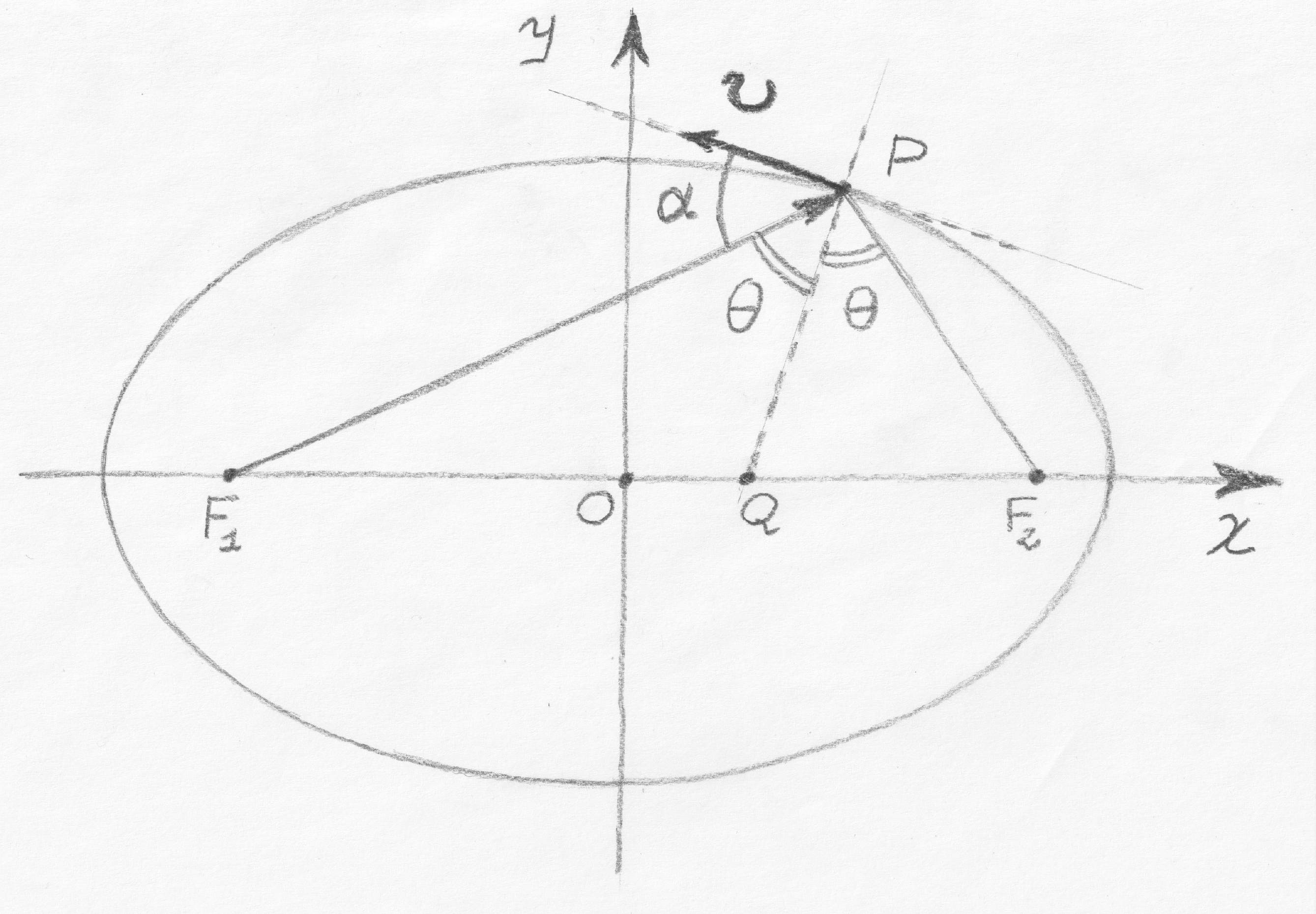}}
\caption{The second Kepler law.} 
\label{Fig:Geometry.1.1}
\end{figure}
With reference to Fig.~\ref{Fig:Geometry.1.1}, write  the areal speed by using the geometrical definition
of cross product, i.e.,
\begin{equation}
\label{Eq:ArealSpeed}
2\dot{\mathcal{A}}\,=\,
\overline{F_1P}\,v\,\sin\alpha\,=\,
\overline{F_1P}\,v\,\cos\theta\,,
\end{equation}
where, due to the property P1, $\theta\,=\,\sphericalangle F_1PQ\,=\,\sphericalangle F_2PQ$.
On applying cosine's law to both triangles  $F_1PQ$ and $F_2PQ$ we have
\begin{equation}
\label{Eq:ArealSpeed.2}
\left\{
\begin{array}{l}
\overline{F_1 Q}^2\,=\,\overline{F_1 P}^2\,+\,\overline{P Q}^2\,-\,2\,\overline{F_1 P}\,\overline{PQ}\,\cos\theta\,,\\
\\
\overline{F_2 Q}^2\,=\,\overline{F_2 P}^2\,+\,\overline{P Q}^2\,-\,2\,\overline{F_2 P}\,\overline{PQ}\,\cos\theta\,,
\end{array}
\right.
\end{equation}
which, on taking Eq.~(\ref{Eq:App:A.2}) into account, after simple algebra gives 
\begin{equation}
\label{Eq:ArealSpeed.2.2}
\cos\theta\,=\,\Dfrac{b}{\left(\overline{F_1 P}\,\overline{F_2 P}\right)^{1/2}}\,.
\end{equation}
On substituting from Eq.~(\ref{Eq:ArealSpeed.2.2}) into Eq.~(\ref{Eq:ArealSpeed}),
the following  relationship between the point speed $v$ and the areal speed $\dot{\mathcal{A}}$
is then obtained:
\begin{equation}
\label{Eq:ArealSpeedAndSpeed}
v\,=\,\Dfrac 2{b}\,\dot{\mathcal{A}}\,
\sqrt{\Dfrac{\overline{F_2 P}}{\overline{F_1 P}}}\,.
\end{equation}
Moreover, the constant value of the areal speed, say $K$, is 
obtained simply by dividing the ellipse area, $\pi a b$, by  the orbital period, say $T$, i.e.,
\begin{equation}
\label{Eq:ArealSpeedAndSpeed.2}
{\dot{\mathcal{A}}}\,=\,K\,=\,\Dfrac {\pi ab}T\,.
\end{equation}

\subsection{Step 3: use of third Kepler's law and finalization of the proof}
\label{Subsec:KeplerLawIII}

We are now ready to finalize the proof, i.e., to show that the modulus of the acceleration
is proportional to the inverse square of $\overline{F_1P}$.
Since  $\acceleration$ is parallel to $\overrightarrow{F_1P}$, the normal component of the acceleration,
say $a_\nu$, is given by $a_\nu=|\acceleration|\,\cos\theta$, with  $\theta=\sphericalangle F_1PQ$
(see again Fig.~\ref{Fig:Geometry.1.1}).
Then, on introducing the curvature radius, say $\rho_P$, of the ellipse at point $P$, we have
\begin{equation}
\label{Eq:NormalAcceleration}
a_\nu\,=\,\Dfrac {v^2}{\rho_P}\,\Longrightarrow\,|\acceleration|\,=\,\Dfrac{v^2}{\rho_P\,\cos\theta}\,.
\end{equation}
In order to continue, the explicit expression of the curvature radius $\rho_P$ is needed.
This is just the second property of ellipses mentioned at the beginning of the paper,
which states:
\begin{quotation}
{\em P2: the radius of curvature of  the ellipse in $P$ is
\begin{equation}
\label{Eq:RaggioCurvaturaEllisse}
\rho_P\,=\,\Dfrac {\left(\overline{F_1P}\,\overline{F_2P}\right)^{3/2}}{ab}
\end{equation}
 }
\end{quotation}
An elementary proof of P2 is outlined in~Appendix~\ref{App:B} and could constitute the subject of
another preliminary practice session.

On substituting from Eqs.~(\ref{Eq:ArealSpeed.2.2}),~(\ref{Eq:ArealSpeedAndSpeed}),~(\ref{Eq:ArealSpeedAndSpeed.2}),
and~(\ref{Eq:RaggioCurvaturaEllisse}) into Eq.~(\ref{Eq:NormalAcceleration}),
we thus have 
\begin{equation}
\label{Eq:NormalAcceleration.2}
\begin{array}{l}
|\acceleration|\,=\,
\Dfrac 4{\cancel{b^2}}\,\Dfrac{\pi^2\,a^2\,\cancel{b^2}}{T^2}K^2\,
\Dfrac{\cancel{\overline{F_2P}}}{\overline{F_1P}}\,
\Dfrac{a\,\cancel{b}}{\overline{F_1P}^{3/2}\,\cancel{\overline{F_2P}^{3/2}}}\,
\Dfrac{\overline{F_1P}^{3/2}\,\cancel{\overline{F_2P}^{1/2}}}{\cancel{b}}\,=\,\\
\\
\,=\,
4\pi\,\left(\Dfrac{a^3}{T^2}\right)\,\,\Dfrac{1}{\overline{F_1P}^2}\,,
\end{array}
\end{equation}
and, finally,  the last step:  Kepler's law III asserts that the ratio 
$a^3/T^2$ must be a constant independent of the planet. 
On denoting $C$ such a constant, Eq.~(\ref{Eq:NormalAcceleration.2}) finally gives
\begin{equation}
\label{Eq:NormalAcceleration.3}
\hspace*{-2cm}
|\acceleration|\,=\,
\Dfrac{4\pi C}{\overline{F_1P}^2}\,,
\end{equation}
\emph{Quod Erat Demonstrandum}.

\section*{Acknowledgment}

I wish to thank Turi Maria Spinozzi for his invaluable help during the preparation of the manuscript.

\appendix

\section{Proof of P1}
\label{App:A}

The starting point is Eq.~(\ref{Eq:EquazioneCartesianaEllisse}), where
both coordinates should be thought of as functions of time $t$ in order to 
describe an hypothetical motion of $P$ on the ellipse. Then, on deriving both sides
with respect $t$ we have
\begin{equation}
\label{Eq:App:A.1}
\Dfrac{x\dot{x}}{a^2}\,+\,
\Dfrac{y\dot{y}}{b^2}\,=\,0\,,
\end{equation}
whose left side  can be interpreted as the scalar product between the velocity of the point,
whose Cartesian representation is $(\dot{x},\dot{y})$,
and the vector $(x/a^2,y/b^2)$. Since such scalar product is null, 
 the latter vector must be directed along the ellipse normal at $P$. 

\noindent On denoting $Q$ the intersection of such normal with the $x$-axis 
(see again Fig.~\ref{Fig:Geometry.1.1}), 
and on taking into account that $\overline{PP'}=y$, it is not difficult to show  that
$Q\equiv (x\,\epsilon^2,0)$. 
Moreover, on taking into account that   
$\overline{F_1 P}=a\,+\,x\,\epsilon$ and  that $\overline{F_2 P}=a\,-\,x\,\epsilon$,\footnote{Proving these relationships could be left to students as a useful geometry problem.}  it follows at once that:
\begin{equation}
\label{Eq:App:A.2}
\begin{array}{l}
\overline{F_1Q}\,=\,\epsilon\,\overline{F_1P}\,,\\
\\
\overline{F_2Q}\,=\,\epsilon\,\overline{F_2P}\,.
\end{array}
\end{equation}
Finally, on applying the
sine theorem to the triangles $PQF_1$ and $PQF_2$,
property P1 then follows.

\section{Proof of P2}
\label{App:B}

To evaluate the radius of curvature, imagine a point moving across the ellipse 
according to the  following law of motion:
\begin{equation}
\label{Eq:App:B.1}
\left\{
\begin{array}{l}
x(t)\,=\,a\,\cos t\,,\\
\\
y(t)\,=\,b\,\sin t\,,
\end{array}
\right.
\end{equation}
with $t\in[0,2\pi]$ in suitable units.
On deriving  both $x$ and $y$,  Cartesian representations for
both velocity and acceleration reads
\begin{equation}
\label{Eq:App:B.2}
\velocity\,=\,\left\{
\begin{array}{l}
\dot{x}(t)\,=\,-a\,\sin t\,,\\
\\
\dot{y}(t)\,=\,b\,\cos t\,,
\end{array}
\right.
\end{equation}
and
\begin{equation}
\label{Eq:App:B.3}
\acceleration\,=\,\left\{
\begin{array}{l}
\ddot{x}(t)\,=\,-a\,\cos t\,,\\
\\
\ddot{y}(t)\,=\,-b\,\sin t\,,
\end{array}
\right.
\end{equation}
respectively.
Then, the radius of curvature stems from the well known intrinsic expression,
\begin{equation}
\label{Eq:App:B.4}
\rho_P\,=\,\Dfrac {v^3}{|\velocity\,\times\,\acceleration|}\,,
\end{equation}
which, on taking Eqs.~(\ref{Eq:App:B.3}) and~(\ref{Eq:App:B.4}) into account,
after simple algebra gives at once
\begin{equation}
\label{Eq:App:B.5}
\hspace*{-2cm}
\begin{array}{l}
\rho_P\,=\,
\Dfrac {(a^2\,\sin^2t\,+\,b^2\,\cos^2t)^{3/2}}{ab}\,=\
\Dfrac {(a^2\,-(a^2\,-\,b^2)\,\cos^2t)^{3/2}}{ab}\,=\,\\
\\
\,=\,
\Dfrac {(a^2\,-\,a^2\,(1\,-\,b^2/a^2)\,\cos^2t)^{3/2}}{ab}\,=\,
\Dfrac {(a^2\,-\,x^2\,\epsilon^2)^{3/2}}{ab}\,=\,\\
\\
\,=\,
\Dfrac {(a\,+\,x\,\epsilon)^{3/2}\,(a\,+\,x\,\epsilon)^{3/2}}{ab}\,=\,
\Dfrac {\overline{F_1P}^{3/2}\,\overline{F_2P}^{3/2}}{ab}\,.
\end{array}
\end{equation}

\end{document}